# Zero-Quantum filtered Perfect NOESY: Enhanced sensitivity for NOE cross peaks perturbed by Zero Quantum artefacts

Bikash Baishya[a]*

[a]Center of Biomedical Research (Formerly Centre of Biomedical Magnetic Resonance), SGPGIMS Campus, Raebareli Road, Lucknow, 226014, India

**Abstract**

Evolution of homonuclear *J*-coupling during $t_1$ period creates Zero Quantum artefacts which seriously degrade the performance of 2D NOESY. In our previous work [28], it is demonstrated that creation of such artefacts can be prevented by incorporation of a perfect echo based broadband homo-nuclear decoupling technique during $t_1$ period that also allows chemical shift encoding. However, such experiments have limited $F_1$ resolution as the perfect echo based decoupling remains efficient for interpulse delay $2\tau$ (=$t_1$) short compared to $1/J$. A solution to this problem is provided in the present work where we show that even a partial decoupling effect of a perfect echo block for extended $t_1$ period close to $1/2J$ greatly suppress the build up of antiphase content and thereby enhances the sensitivity of corresponding NOE cross peaks. The partial decoupling do lead to generation of partial ZQ coherence which is further eliminated by application of a ZQ filter.

**Introduction:**

Achieving Homonuclear $^1$H-$^1$H or $^{13}$C-$^{13}$C scalar decoupling is an important concern in 2D NMR. Replacing the multiplets with singlets leads to improved resolution, gain in sensitivity, and suppression of certain artefact. Two important approaches for achieving this objective are: constant time evolution can remove the effect of *J*-coupling for the indirect $t_1$ dimension [1-3]; and in case of heteronuclear experiments application of a single BIRD [4] block eliminates such unwanted effects of *J*-modulation. Other noteworthy approaches are use of spatially-selective $^1$H spin inversions in Zangger/Sterk sequence [5-7], and utilization of selective pulses whenever feasible [8-9]. The Zangger/Sterk or ZS method suffers from low intrinsic sensitivity and other methods that rely on selective pulses are applicable to the study of only limited resonances and favourable in situation when applications of selective pulses are feasible. The perfect echo based homodecoupling has also been utilized to many advantages in many recent publications [10-14].

An undesired consequence of homonuclear *J*-evolution in indirect dimension of NOESY is the Zero-Quantum artefact. The NOESY experiment has played an outstanding role in structure elucidation of molecules in solution. The NOE cross peaks between protons provides information on through space distance between them and arises from magnetization transfer via dipole-dipole cross relaxation during the mixing period [15-16]. The magnetization must be along the z-axis for such transfer during the mixing time. Only z-magnetization should contribute to the NOE cross peaks, and either phase cycling or gradient is employed to ensure that. However, Zero Quantum (ZQ) coherence cannot be discriminated from z-magnetization by such phase cycling or gradient. The homonuclear *J*-evolution during $t_1$ period creates the antiphase states which subsequently coverted into ZQ and other higher quantum coherences representing the magnetization loss and the artifacts created that contaminated the desired NOE cross peaks.

The ZQ peaks also called ZQ artefacts have antiphase lineshape with long dispersive tail. Utilization of longer mixing time can reduce the intensity of ZQ coherence, but, spin diffusion contribution starts dominating the NOE cross peaks under such conditions leading to incorrect molecular structure. Spin diffusion is minimized at short mixing time, but, such experiments display weaker NOE cross peak intensities overlapped by stronger ZQ peaks. The intensity of a cross peak has strong dependence on the distance between protons involved in that cross peak. Identical frequencies of NOE and ZQ peaks in $F_1$ and $F_2$ dimensions, hampers accurate quantification of NOE cross peaks. The long dispersive tails further creates other problems such as error in the integration of other cross peaks, reduce effective resolution and introduce unwanted correlations. A number of methods have been proposed [9, 17-22] to suppress ZQ coherence that is inherent to the pulse scheme employed in NOESY. One such noteworthy recent approach utilizes band selective homonuclear decoupling in both dimensions of NOESY for enhancing sensitivity and suppressing ZQ [9]. The method is elegant for peptides and proteins as it involves band selective excitation of amide and/or H-alpha protons. Another method of pure shift NOESY has been realized by incorporation of ZS method in direct dimension and covariance processing but does not involve an actual $\omega_1$ experimental decoupling [23].

Broadband $\omega_1$ decoupling in 2D NOE has also been carried out by time reversal of the evolution under scalar spin-spin interactions [24]. This experiment is without the disadvantages of constant time experiments but pays sensitivity penalty by a factor of 2.5 in comparison to conventional NOESY. This method also produces some unwanted cross peaks. Other approaches are broadband proton decoupled proton spectra obtained by 45° degree projection of the diagonal-peak multiplets of an anti z-COSY spectrum [25].

Very recently, we have demonstrated a novel modification of original perfect echo [26-27] based broadband homonuclear decoupling that allows chemical shift encoding as well as homo-decoupling for limited $t_1$ period in NOESY for preventing the creation of Zero Quantum artefact [28]. In this approach, removal of the $2^{nd}$ refocusing pulse of the double spin echo involved in the conventional perfect echo allows one to encode chemical shift without compromising the *J*-refocusing property as long as inter pulse delay $t_1/4$ is smaller than $1/J$. The limitations of the sequence is that for longer $t_1$ period, partial zero quantum appears along with NOE cross peaks due to partial refocusing of homonuclear *J*-coupling interaction.

Herein, it is demonstrated that even a partial decoupling effect of a perfect echo block for longer $t_1$ period greatly reduces the build up of antiphase content and thereby enhances the sensitivity of the corresponding NOE cross peaks. The partial decoupling for longer $t_1$ period of 2D PE-NOESY do generate partial ZQ which is eliminated by application of a ZQ filter. The method is very simple in implementation and enhances the sensitivity for NOE cross peaks without any interference from Zero-Quantum artefacts and without extending experimental time.

**Description of the 'perfect echo' ZQ-filtered NOESY pulse sequence:**

The perfect echo based ZQ-filtered NOESY sequence is shown in Fig. 1A and a conventional ZQ filtered NOESY in Fig. 1B. The ZQ filtered perfect echo NOESY differs from conventional ZQ filtered NOESY by removal of the 1st 90˚ pulse and the $t_1$ evolution period of the conventional ZQ filtered NOESY by the modified PE scheme shown as dotted box from 'a' to 'f' in Fig. 1B. The modified PE scheme removes the 2nd refocusing pulse (shown as dotted pulse at time point 'e') of conventional PE scheme so that chemical shift can evolve between time point 'd' and 'f' before the magnetization is stored along the z-axis by the 90˚ pulse after time point 'f'. The 1st part of the modified PE allows *J*-prefocusing from time point 'b' to 'c' while the *J*-refocusing 90˚ pulse between time point 'c' and 'd' exchanges the antiphase magnetization between coupled spins. Subsequently *J*-coupling refocuses at time point 'f'. All the 'd0' delays in the modified PE bloc pertain to $t_1/4$ and are incremented simultaneously. The *J*-prefocusing and refocusing requires the $t_1$ increments in PE-NOESY to be two times longer than that in a conventional ZQ filtered NOESY. The chemical shift is encoded only during half of each $t_1$ increments i.e. for the time interval between 'd' and 'f' (which is equal to that from time point 'b' to 'c' in Fig. 1A) so that maximum chemical shift evolution is identical in both the sequences. Owing to two times longer $t_1$ increments in PE-NOESY, the $T_2$ decay is also twice than that of conventional NOESY but experimentally that extra loss of magnetization could be compensated by *J*-refocusing mediated signal gain which could otherwise be converted to ZQ and other unwanted coherences. The open box with a diagonal line during the mixing period from 'd' to 'g', represents a swept-frequency 180 pulse of duration $\tau_f$ applied in conjunction with a small gradient $G_1$ and acts as the ZQ filter.

The broadband homodecoupling is efficient as long as $t_1/4 \ll 1/J_{HH}$. However, $J_{HH}$ is quite small (5-15 Hz), and therefore a $t_1^{max}/4$ value of up to 10 ms is considered efficient for homodecoupling and efficient suppression of ZQs were observed in our previous work [28]. The requirement, $t_1^{max}/4 \ll 1/J_{HH}$ in PE-NOESY implies that the value of $t_1^{max}$ should not be high; otherwise, only partial suppression of ZQs will be observed once $t_1^{max}$ exceeds certain value. However, a smaller $t_1^{max}$ limits the resolution in PE-NOESY spectra and a longer $t_1^{max}$ generates partial ZQ peaks. In our earlier work we have been able to encode chemical shift for a $t_1^{max}$ of 20ms along with an efficient suppression of ZQs. However, as $t_1^{max}$ approaches 1/2J, the antiphase content becomes maximum with negligible in phase state. This significantly reduces the z magnetization necessary for NOESY transfer. Consequently, loss of magnetization to ZQ and other higher quantum terms through *J*-antiphase states becomes significant for such longer $t_1$ values. And even a partial *J*-refocusing for such longer $t_1$ increments by perfect echo based ZQ filtered NOESY, can significant enhance the sensitivity provided the partial ZQ suppression is ensured by implementation of a ZQ filter during $t_1$.

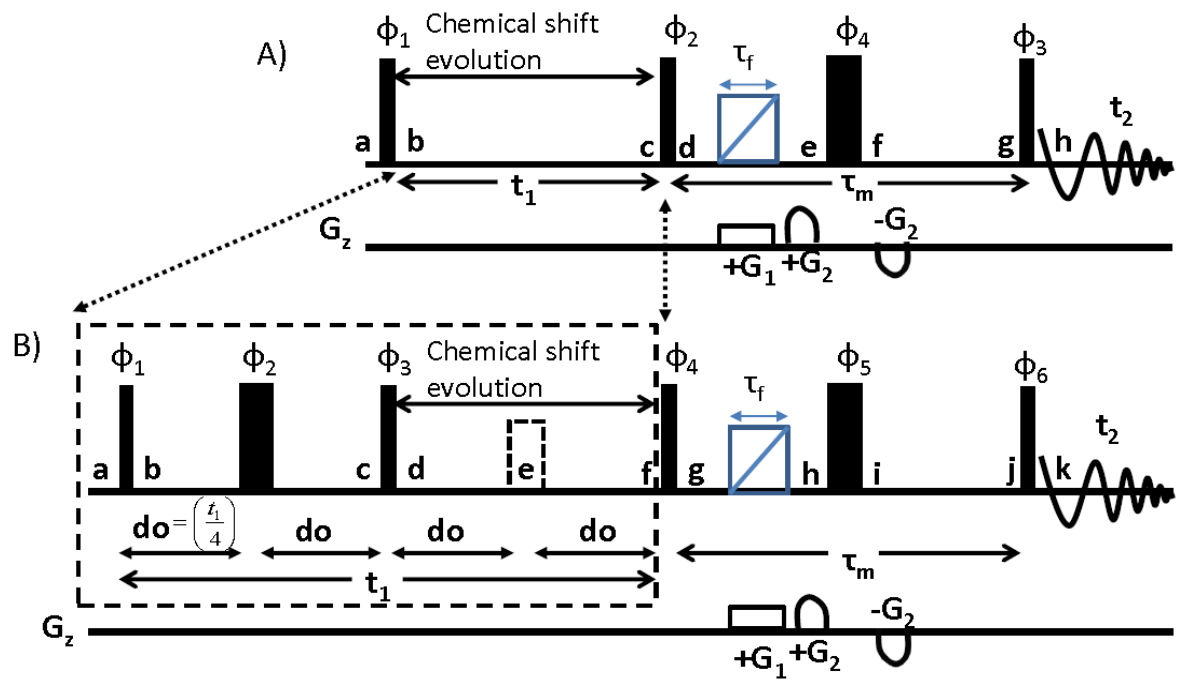

**Fig. 1(A)** A conventional ZQ filtered 2D NOESY pulse sequence. The open box with a diagonal line during the mixing period from 'd' to 'g', represents a swept-frequency 180 pulse of duration $\tau_f$ applied in conjunction with a small gradient $G_1$. This acts as the ZQ-filter. The phases are $\Phi_1$=02, $\Phi_2$=8(0), $\Phi_3$=00221133, $\Phi_4$=0 and $\Phi_R$=02201331. **(B)** ZQ filtered perfect echo 2D NOESY pulse sequence, the phases are $\Phi_1$=02, $\Phi_2$=8(0), $\Phi_3$=4(0)4(2), $\Phi_4$=0, $\Phi_5$=1133, $\Phi_6$=1313 and $\Phi_R$=02202002. The dotted box from time point 'a' to 'f' is a modified perfect echo block for converting the conventional NOESY sequence into Perfect Echo NOESY. The modified PE scheme removes the 2nd refocusing pulse (shown as dotted pulse at time point 'e') of conventional PE scheme so that chemical shift can evolve between time point 'd' and 'f' before the magnetization is stored along the z-axis by the 90˚ pulse after time point 'f'. The 1st part of the modified PE allows *J*-prefocusing from time point 'b' to 'c'. Application of the 90˚ pulse at 'c' refocuses *J*-coupling at time point 'f'. All the 'd0' delays in the modified PE bloc pertain to $t_1$/4 and are incremented simultaneously. The *J*-prefocusing and refocusing requires the $t_1$ increments in PE-NOESY to be two times longer than that in a conventional NOESY. The chemical shift is encoded only during half of each $t_1$ increments i.e. for the time interval between 'd' and 'f' which is equal to that from time point 'b' to 'c' in Fig. 1(b). All gradients are along z axis. $G_1$=4, $G_2$:$G_3$= 30:-30. States TPPI is used for frequency discrimination in $t_1$. ZQ filter is applied as in (A).

**Experimentals:**

For experimental demonstration of the sensitivity enhancement of cross peaks in ZQ filtered PE-NOESY [Fig. 1(B)] over conventional ZQ filtered NOESY [Fig. 1(A)], these experiments were performed on Cyclosporine A– a cyclic undecapeptide (50mM in deuterated benzene, $C_6D_6$) on a Bruker avance 800 MHz NMR spectrometer equipped with a cryoprobe at 288 Kelvin. ZQ suppression was achieved by Killer-Tripleton filter. Details of experimental and processing parameters are given in figure captions.

**Results and discussions:**

The perfect echo based NOESY experiments relies on refocusing of homonuclear *J*-refocusing during $t_1$ dimension to prevent build up of ZQ coherence and other higher order coherences. Nevertheless, we observed that for longer $t_1^{max}$, the perfect echo achieves partial refocusing leading to the appearance of ZQ in the PE-NOESY spectra. However, the amplitude of these ZQs are less than that in a conventional NOESY without any ZQ filter. This indicates that if the partial ZQ is eliminated in PE-NOESY, then the NOE cross peaks will show an elevated intensity relative to conventional ZQ filtered NOESY. In addition, this will allow this experiment to be carried out in a high resolution mode in the indirect dimension. To systematically investigate these advantages, the ZQ filtered PE-NOESY and conventional ZQ filtered NOESY sequence as shown in Fig 1A and 1B were recorded for comparison on Cyclosporine A. A few NOE cross peaks from these experiments are displayed in panels 1-12 in Fig 2 for comparison. In each panel, the left blue colour peak corresponds to ZQ filtered PE-NOESY while the right side red colour peak is from conventional ZQ filtered NOESY plotted on same contour levels. A short mixing time of 250 ms and a $t_1^{max}$ of 50 ms was utilized to ensure a strong interference between ZQ coherence and NOE in both version of Fig. 1. For ZQ suppression, adiabatic smoothed CHIRP swept-frequency 180° pulses were used sweeping through 26 kHz in 16 ms with field strength of 1.1 kHz. The $t_1^{max}$ of 50 ms utilized in these experiments is considered close to 1/2J and hence significant loss of magnetization as ZQ and other higher quantum coherence (through antiphase states) is expected in conventional ZQ filtered NOESY compared to ZQ filtered PE-NOESY.

In all panels 1-12, ZQ filtered PE-NOESY reveals cross peaks (blue colour) free from ZQ artefact and with better sensitivity compared to conventional ZQ-filtered NOESY (red colour). The homo-decoupling efficiency of perfect echo declines with longer $t_1^{max}$ value, nevertheless, even this partial decoupling suppress creation of unwanted coherences and hence considered valuable for raising the sensitivity of cross peaks. As $t_1^{max}$ approaches 1/2J value, the antiphase terms approaches its maximum value and even a partial *J*-refocusing can greatly reduce the antiphase content while raising inphase magnetization required for sensitivity enhancement. However, this enhanced sensitivity would not be observable if the partial ZQs which eclipse NOE cross peaks are not eliminated. This is achieved in the ZQ filtered PE-NOESY. Suppression of entire ZQ peak is noteworthy in ZQ filtered PE-NOESY and ZQ filtered NOESY in the blue and red coloured peaks in all the panels respectively. For short mixing time the NOE build up is quite low whereas ZQ amplitude is quite high. However, once the suppression of ZQ is achieved in both sequences, the NOE peaks can be observed with more convenience and relative intensities of these peaks can be compared in both approaches.

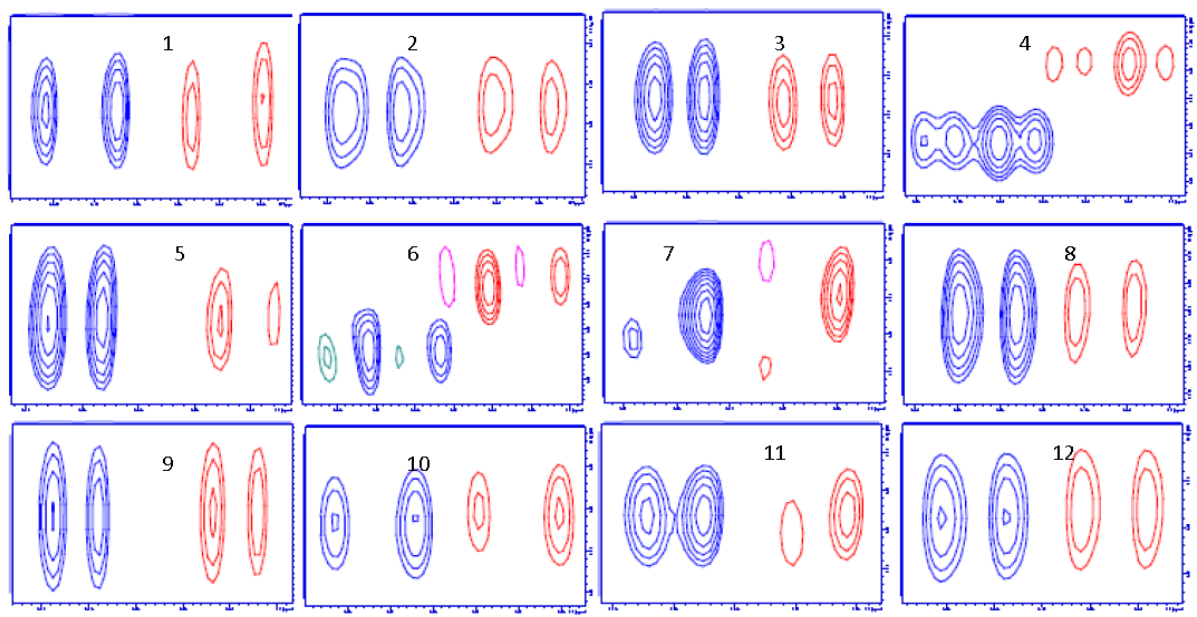

**Fig. 2.** Panels 1-12 displays a few NOESY cross peaks of Cyclosporine A (50 mM in deuterated benzene) at 800 MHz $^1$H frequency, 15° C. In each panel, the left peak represents ZQ filtered PE-NOESY while the right peak represents conventional ZQ filtered NOESY plotted on same contour levels. A short mixing time of 250 ms was used so that NOE intensity is low whereas ZQ amplitude is high. The 2D data is recorded with acquisition times of 250ms ($t_2$) and 50ms ($t_1$) respectively, 8 scans for each $t_1$ increment with relaxation delay of 1.35 sec. For ZQ suppression, adiabatic smoothed CHIRP swept-frequency 180° pulses were used sweeping through 26 kHz in 16 ms with field strength of 1.1 kHz. No ZQ peak is observed in any of the panels. The enhanced sensitivity of the ZQ filtered PE-NOESY cross peaks (left blue colours) are clearly visible relative to their conventional ZQ filtered NOESY counterpart (right red colours) even for a longer $t_1$ acquisition time of 50 ms. Thus, the partial homonuclear *J*-refocusing effect of perfect echo even for higher $t_1^{max}$ value recovers a good fraction of magnetization in PE-NOESY which is lost as ZQ as well as other higher quantum coherences in conventional NOESY.

**Conclusion:**

In conclusion, it is demonstrated that perfect echo based broadband homonuclear decoupling along with a ZQ filter can be carried out even for longer $t_1$ period of 2D NOESY.

We have demonstrated a $t_1^{max}$ of 50 ms in this work, which is close to $1/2J_{HH}$. The partial decoupling for longer $t_1$ period results in NOESY spectra that displays NOE cross peaks with better sensitivity than that is achieved in a conventional ZQ filtered NOESY. Since the quality of structures determined is directly dependent on the number of detectable NOEs, raising the cross peak intensities is an important concern. Although the $T_2$ decay of magnetization in $t_1$ period of ZQ filtered PE-NOESY is twice than that of conventional ZQ filtered NOESY, the SNR gain in the former relative to the later implies that loss of magnetization via *J*-evolution is more severe for longer $t_1$ period of NOESY. Thus the present work represents a significant improvement as it improves sensitivity of 2D NOESY with a broadband homo-decoupling approach. Similar applications are possible in other 2D correlation experiments and will be published elsewhere.